\documentclass[aps,twocolumn,showpacs]{revtex4}
\usepackage{graphicx}
\begin{document}
\title{Mechanically induced current and quantum evaporation from
Luttinger liquids}
\author{Andrei Komnik and Alexander O. Gogolin}
\affiliation{Department of Mathematics, Imperial College, 180 Queen's Gate,
London SW7 2BZ, United Kingdom}
\date{\today}
\begin{abstract}
We investigate transport through a tunnelling junction between an
uncorrelated metallic lead and a Luttinger liquid when the latter
is subjected to a time dependent perturbation. The tunnelling
current as well as the electron energy distribution function are
found to be strongly affected by the perturbation due to
generation of harmonics in the density oscillations. Using a
semiconducting lead instead of a metallic one results in electrons
being injected into the lead even without applied voltage. Some
applications to carbon nanotubes are discussed.
\end{abstract}

\maketitle
\section{Introduction} \label{introduction}
The influence of static impurities on the transport properties of
one-dimensional (1D) correlated electron systems is well
understood \cite{kanefisher,furusakinagaosa} (for a recent review
on 1D physics see e.~g.~\cite{book}). On the contrary, the effect
of dynamic time-dependent perturbations on transport from open
Luttinger liquids (LLs) has been barely addressed in the
literature. Yet, investigations into the role of dynamic sources
in other correlated systems reveal interesting phenomena. So, it
was seen in quantum evaporation (QE) experiments from superfluid
helium \cite{qenature} that such perturbations can significantly
influence the structure of the complicated collective particle
state. A phonon source (which is essentially a dynamic
perturbation), imbedded into a superfluid is able to excite single
particles above the energy needed to escape from the condensate.
In this paper we address the question whether a correlated 1D
electron system with open boundary shows similar physics. Research
in this direction is particularly interesting because of a
possibility to make related measurements on single-wall carbon
nanotubes (SWNTs). Their electronic degrees of freedom are known
to be described by the (four-channel) LL model
\cite{bockrath1,sammlung1,sammlung3,dekker}.

An experimental realisation of a periodic perturbation can, for
example, be achieved by laser radiation which causes localised
oscillations of the lattice coupled to the electron density.
Alternatively, it has been realised by the oscillation of the
nanotube itself \cite{deheer}. If the transport through the
nanotube changes significantly in the presence of such perturbation,
it could be used to gain information about the cause of the
oscillation. This would add to already known applications of
nanotubes as detector or sensor devices \cite{actuator}.

The system under investigation is a half-infinite LL with an open
end. At the boundary it is contacted by an electrode which can be
metallic, semiconducting or is even opened into vacuum (the latter
case can be modelled by a metallic electrode with the chemical
potential sent to infinity as in case of the field emission effect
\cite{plummer,my}). We are interested in the current-voltage
characteristics and in the electron energy distribution function
(also called the total energy distribution function: TED) in the
presence of a dynamic scatterer inside the LL. The TED is
proportional to the energy resolved current in the case of
emission into vacuum.

In the experimental setups similar to that of Ref.~\cite{deheer},
the nanotube oscillates as a whole. Clearly such oscillations
strongly affect the phase of the electron wave-function, but cause
little elastic stress and hence only a weak back-scattering. It
turns out, somewhat surprisingly, that a dynamic forward
scattering perturbation non-trivially effects the electron
transport. Contrary to a back-scattering perturbation,
\emph{static} forward scattering contributes neither to the TED
nor to the current, so that the contribution of the dynamic part
can be singled out. Because of these reasons, we concentrate here
on the forward scattering model, which is very simple and exactly
solvable and yet leads to interesting predictions for the electron
transport.

The paper is organised as follows. In the next Section we discuss
the TED of electrons in the vicinity of the open end in the presence
of a time-dependent perturbation. We derive a general formula for
the TED. In the
subsequent Section we derive a similar formula for the tunnelling
current across the interface to an additional electrode. Section
\ref{sectionapplications} is then devoted to applications of the
results to different situations and contains discussion of the
main findings.

\section{Time-dependent perturbation in 1D} \label{sectionTED}
\begin{figure}
\includegraphics[scale=0.35]{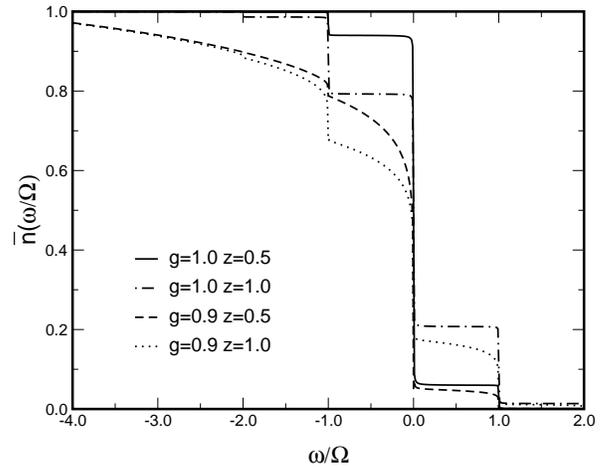}
\caption[]{\label{1stplot} The TED for a non-interacting system
($g=1$) and for a weakly interacting LL ($g=0.9$) at different
oscillator strengths. The Fermi energy $E_F$ is set to zero.}
\end{figure}

Let us begin by recollecting some known results for the TED of
electrons in the presence of a time-dependent perturbation. Since
a static forward scatterer only changes the phase of the wave
function, it cannot influence the TED. As soon as the perturbation
becomes time-dependent, the scattering process is not elastic any
more and particles can acquire or lose energy. Hence the TED is
changed. In the case of a periodic perturbation with the frequency
$\Omega$, the electron can lose or gain the energy $n\Omega$ upon
scattering, $n$ being an integer. Thereby an infinite number of
equidistant sidebands emerge in the TED even in non-interacting
systems \cite{hauge,bagwell}. As a result, the TED is distorted in
a staircase manner as shown in Fig.~\ref{1stplot}. We stress that
the forward scattering is sufficient to cause these effects.

This can be quantified for our system using the open boundary
formalism of Ref.~\cite{fg}.
According to \cite{fg}, in an open system the electrons can be regarded as
chiral particles living in an infinite system.
Thereby the operator $\psi(x,t)$
on the negative half-axis describes the electrons moving towards
the boundary (right-movers) and for positive $x$ it
corresponds to electrons moving in the opposite direction
(left-movers). Since we are dealing with an explicitly
time-dependent situation the TED $n(\omega,t)$ at the boundary should
be defined using the Wigner representation,
\begin{equation} \label{TEDdefinition}
 n(\omega,t) = \int \, d \tau \, e^{i \omega \tau} \langle
 \psi^\dag(0,t-\tau/2) \psi(0,t+\tau/2) \rangle \, .
\end{equation}
Because of the open boundary condition imposed on the electron
wave functions the probability density to find any particles
\emph{exactly} at the boundary is, of course, zero. However, we assume
the TED to be measured on the second last site of the
underlying lattice model so that $x$ in the last formula is
actually not $0$ but the lattice constant $a_0$.

Since we are dealing with a non-equilibrium situation, we
resort to the Keldysh formalism \cite{keldysh,LLX}, in which the TED
can be conveniently expressed through one of the off-diagonal
Green's functions (from now on we drop the spacial coordinate),
\begin{equation}
 n(\omega,t) = -i \int \, d \tau \, e^{i \omega \tau}
 g^{-+}(t+\tau/2,t-\tau/2) \, ,
\end{equation}
where (we give all Green's functions for future reference),
\begin{eqnarray}                                 \label{Gfungendefine}
 g^{-+}(t,t') &=& i \langle \psi^\dag(t') \psi(t) \rangle \, , \nonumber \\
 g^{+-}(t,t') &=& -i \langle \psi(t) \psi^\dag(t') \rangle \, , \nonumber \\
 g^{--}(t,t') &=& -i \langle T \psi(t) \psi^\dag(t') \rangle \, , \nonumber \\
 g^{++}(t,t') &=& -i \langle \widetilde{T} \psi(t) \psi^\dag(t') \rangle \,
 .
\end{eqnarray}
Here $T$ and $\widetilde{T}$ denote the time- and
anti-time-ordering operations, respectively. Notice that these
functions are not translationally invariant in time because of the
explicit time dependence of the perturbation $U(t)$. All
calculations are most transparent in the bosonized representation
usually used in the theory of LLs \cite{haldane}. The interacting
electron field $\psi(t)$ can then be expressed in terms of a free
Bose field $\phi(t)$. The physical interpretation being that
$\phi$ describes the collective low-energy plasmon excitations in
our correlated electron system. At the system's boundary we have
\cite{fg}:
\begin{eqnarray} \label{psidef}
 \psi(t) = \frac{1}{\sqrt{2 \pi a_0}} e^{i \phi(t) / \sqrt{g}} \, ,
\end{eqnarray}
This equation contains the Luttinger liquid parameter $g$,
which is related to the interaction strength $U_0$
via  $g=(1+ 4 U_0/\pi)^{-1/2}$ \cite{haldane}.
The mode expansion for the field $\phi(x)$, in terms of
boson creation $b^\dag_q$ and annihilation $b_q$ operators, reads
\begin{eqnarray}                                    \label{phidef}
 \phi(x) = \sum_{q>0} \sqrt{\frac{\pi}{q L}}
\Big[ e^{i q x} b_q + e^{-i q x} b_q^\dag \Big] e^{-a_0 q/2} \, .
\end{eqnarray}
We assume the system having length $L$, so that the momentum $q$
takes quantised values $q=2 \pi n/L$, where $n$ is an integer. At
the end of all calculations we shall send $L$ to infinity (any
zero-modes omitted in Eq.~(\ref{phidef}) will drop out in this
limit). In this representation, the (perturbed) LL Hamiltonian
takes the following form,
\begin{eqnarray}  \label{principalH}
 H &=& H_{LL}[\psi] + U(t) = v \sum_{q>0} q b_q^\dag b_q \nonumber \\
 &+&  \int \, dx \, [ \Delta(x,t) + \Delta(-x,t)] \rho(x) \, ,
\end{eqnarray}
where $\Delta(x,t)$ is the forward-scattering time-dependent perturbation and
$v=v_F/g$ is the renormalised sound velocity, $v_F$ is the Fermi
velocity of the non-interacting system and $\rho(x)$ is the
particle density operator. Notice that we are already working in
the chiral representation therefore the last integral is performed
over the whole real axis. For the same reason the scattering
potential appears in the symmetrised form. In terms of the
Bose field, the electron density $\rho(x)$ reads
\begin{eqnarray}                                 \label{density}
 \rho(x) = \psi^\dag(x)\psi(x) = \frac{k_F}{\pi} + \frac{1}{2 \pi \sqrt{g}}
\partial_x \phi(x)
 \, .
\end{eqnarray}
The constant contribution (average density) gives rise to an
overall energy shift and will therefore be dropped in what
follows. Any additional terms containing back-scattering processes
are dropped as well. Combining this relation with
Eq.~(\ref{phidef})
we rewrite the perturbation in terms of the Bose operators,
\begin{equation}
 U(t) = i \frac{1}{2} \sum_{q>0} \sqrt{\frac{q}{\pi g L}}
 [\Delta_q(t) + \Delta_{-q}(t)](b_q-b^\dag_q) e^{-a_0 q/2} \, ,
\end{equation}
where $\Delta_q(t)$ is the Fourier transform
of the perturbation amplitude, $\Delta_q(t) = \int \, dx \, e^{i q x}
\Delta(x,t)$. To evaluate the Green's functions all we need
is the time dependence of the Bose operators $b_q$.
In our simple model, the latter
can easily be obtained as a solution of the corresponding equation
of motion, which has the following form,
\begin{eqnarray}
 && i \frac{\partial}{\partial t} b^\dag_q(t) = [b^\dag_q, H] =
 \nonumber \\ \nonumber
 &=& -v q b^\dag_q(t) - i \frac{1}{2}\sqrt{\frac{q}{\pi g L}}[\Delta_q(t) +
 \Delta_{-q}(t)] b^\dag_q e^{-a_0 q/2} \, .
\end{eqnarray}
The solution of this equation is
\begin{eqnarray}
 b^\dag_q (t) = (b^\dag_q + f(t))e^{i v q t} \, ,
\end{eqnarray}
where
\begin{eqnarray}
 f(t) = - \frac{1}{2} \sqrt{\frac{q}{\pi g L}}
 \int_{-\infty}^{t} \, d \tau \, [\Delta_q(\tau) + \Delta_{-q}(\tau)]
e^{-i v q \tau- a_0 q/2}  \, . \nonumber
\end{eqnarray}
Using this solution we can write down the Keldysh Green's function
with indices $(ij)$ as products of the bare Green's function
$g_0^{ij}(t-t')$ (i.~e. without the perturbation $U(t)$) and
multipliers, which are responsible for the breaking of the
translational symmetry in time,
\begin{eqnarray}                            \label{factorisation}
 g^{ij}(t,t') = g_0^{ij}(t-t') e^{i(\chi(t)-\chi(t'))} \, ,
\end{eqnarray}
where the phase factors are
\begin{equation}
\chi(t) = - \frac{1}{2 g} \int_0^\infty \, d \tau \,
[ \Delta(-v \tau,t-\tau) + \Delta(v
\tau,t-\tau)] \, .
\end{equation}
Remarkably, factorisation (\ref{factorisation}) holds for interacting systems;
it is a consequence of bosonization.

We are now able to establish the connection between the TED in the
presence of a dynamic scatterer and that of the unperturbed system
$n_0(\omega)$. In order to accomplish this we combine
Eqs.~(\ref{TEDdefinition}), (\ref{Gfungendefine}) and
(\ref{factorisation}). As a result we obtain
\begin{eqnarray}
n(\omega,t) = \int \, \frac{d \omega'}{2 \pi} n_0(\omega')
\int \, d \tau \, e^{i (\omega-\omega') \tau}
\nonumber \\
\times e^{i (\chi(t+\tau/2)-\chi(t-\tau/2))} \, .
\end{eqnarray}

Let us from now on concentrate on a separable harmonic
perturbation, adequate for describing the setup of
Ref.~\cite{deheer}:
\begin{eqnarray} \label{harmonic}
\Delta(x,t) = \sin[ \Omega t] \eta(x) \, .
\end{eqnarray}
We define the function ${\cal F}(\tau)$ responsible for the
geometry of the scatterer,
\begin{equation}
{\cal F}(\tau) = - \frac{1}{2 g} [ \eta(-v \tau) + \eta(v \tau) ]
\, .
\end{equation}
After simple algebra we obtain
\begin{eqnarray} \label{fullnomega}
n(\omega,t) = \int \, \frac{d \omega'}{2 \pi} n_0(\omega')
\int \, d \tau \, e^{i (\omega-\omega') \tau}\nonumber \\
\times \exp \Big( - i \, 2 \, \sin [\Omega \tau/2] \, |z| \, \sin
[\Omega t + \Phi] \Big) \, ,
\end{eqnarray}
where the constant $z$ and the phase delay $\Phi$  between the
TED oscillations and those of the perturbation are
\begin{eqnarray}
 z &=& \int_0^\infty \, d\tau \, e^{i \Omega \tau} {\cal
 F}(\tau) \, , \\
 \Phi &=& \mbox{arcsin} [ \mbox{Re}[z]/|z| ]\, .
\end{eqnarray}
In realistic systems such as SWNTs the frequency $\Omega$ is
expected to be quite high (in the far MHz range) so that it is
unlikely that the explicit time resolution of the TED or other
physical properties of the system would be experimentally
accessible. Therefore it is instructive to investigate
time-averaged quantities (i.~e. averaged over the period of the
perturbation). We define the averaged TED as
\begin{equation}
 \bar n(\omega) = \frac{\Omega}{2 \pi} \int_0^{2 \pi/\Omega} \,
 d t \, n(\omega,t) \, .
\end{equation}
Applying this prescription to Eq.~(\ref{fullnomega}) one finds
that
\begin{eqnarray}
 \bar n(\omega) = \int \, \frac{d \omega'}{2 \pi} n_0(\omega')
 \int \, d \tau \, e^{i (\omega-\omega')
 \tau} \nonumber \\ \times
 J_0[ 2 |z| \sin [\Omega \tau/2]] \, ,
\end{eqnarray}
where $J_0$ is the Bessel function of the first kind. The
$\tau$ integral can be performed by using the formula
\begin{eqnarray} \nonumber
 \int d t \, e^{i \omega t} J_0[ y \sin [\Omega t/2]] = 2 \pi
 \sum_{n=-\infty}^{\infty} \delta\left(\frac{\omega}{\Omega} -
 n \right) J_n^2(y/2) \, .
\end{eqnarray}
As a result one obtains the following expression:
\begin{equation} \label{finalnomega}
 \bar n(\omega) = \sum_{n=-\infty}^{\infty} \, n_0(\omega-n
 \Omega) \,
 J_n^2[|z|] \, ,
\end{equation}
where $z$ is the only free parameter in this formula. It contains
all essential information about the scatterer properties. Hence,
the TED of a system with a harmonic perturbation is a weighted
superposition of infinite number of original unperturbed TEDs
centred at energies $n \Omega$. This result is in accordance with
previous findings for non-interacting systems
\cite{hauge,bagwell}. The important fact is that factorisation
(\ref{factorisation}) continues to hold even for interacting
systems so that formula (\ref{finalnomega}) still applies.

Due to the presence of the oscillating perturbation
the actual TED is distorted even for non-interacting systems, see
Fig.~\ref{1stplot}. For a weak scatterer $|z| \ll 1$ (in the
figure $|z|$ is set to  $0.5$-$1.0$ in order to make the details of
the plot clearer) in a non-interacting system, where $n_0(\omega)$
is just a step function, the particles can barely receive energy
larger than $\Omega$ upon scattering, so that the TED
$\bar{n}(\omega)$ only acquires additional step of the width
$\Omega$ above the Fermi energy $E_F$ and a dip of the same width
just below $E_F$. At higher $|z|$ additional sidebands emerge and
at sufficiently large $|z|$ the TED is bounded from above only by
the actual bandwidth of the host material. The latter can be quite
high and exceed the work function so that particles can leave the
system, or to evaporate from it in analogy to QE.

The TED $n_0(\omega)$ of any unperturbed system is a product of
the Fermi distribution function and the energy-dependent local
density of states (LDOS). Therefore the TED $\bar{n}(\omega)$ for
an LL differs from that for the non-interacting system only by the
energy-dependent details in vicinity of each step. For example, in
Fig.~\ref{1stplot} we plotted the TED for an LL with interaction
parameter $g$, where the LDOS is given by
\begin{eqnarray}          \label{LLDOS}
\nu_\psi(\epsilon) = C_\psi |\epsilon|^{1/g-1} \, .
\end{eqnarray}
$C_\psi$ is a constant, which for a non-interacting system
coincides with the energy independent LDOS \cite{haldane,book}.
The above discussion applies at zero-temperature. It is not
difficult to generalise the formulas to finite temperatures. This
is not of immediate interest though as clearly the effect of finite
temperatures will be to smear out the side-bands. We also note
that the assumption that the perturbation is harmonic,
Eq.~(\ref{harmonic}), is not really restrictive. The same
qualitative behaviour of the TED will persist for any periodic
perturbation.

Formula (\ref{finalnomega}) has important implications for the
field emission theory. The TED of emitted electrons above the
Fermi edge exists only if interactions are present in the emitter
material and shows a divergent behaviour towards $E_F$
\cite{plummer,my}. If, however, we have an additional
time-dependent perturbation, the high-energy tail cannot be
accounted for solely by interactions. In such situation we expect
the field emission (Auger) singularity to emerge at every $n
\Omega$, however, they would be weaker than in the case
without the time-dependent perturbation.

\section{General results for the tunnelling current} \label{sectioncurrent}
We now turn to the calculation of the current-voltage characteristics
of the system which is contacted by an additional electrode at $x=0$. The
corresponding Hamiltonian,
\begin{eqnarray}
H = H_{LL}[\psi] &+& H_0[c]
\\ \nonumber
&+& \gamma \left[ \psi^\dag(0) c(0) + c^\dag(0) \psi(0) \right] + U(t)
\, ,
\end{eqnarray}
contains a contribution $H_0[c]$ describing the electronic
degrees of freedom in the uncorrelated lead (on the right of the
tunnelling junction). Here $\gamma$ is the tunnelling amplitude between
the leads. The current operator can be derived in the usual way
calculating the time derivative of the particle number operator on
the left or, equivalently, on the right side of the junction. Its
average should be calculated by means of the Keldysh technique because of
the explicit time dependence of $U(t)$ and the finite voltage $V$
which is applied to the junction \cite{keldysh,LLX},
\begin{eqnarray} \label{current0}
j(t) = i \gamma \langle [c^\dag(0,t) \psi(0,t) -
\psi^\dag(0,t) c(0,t) ] S_C \rangle_0 \, ,
\end{eqnarray}
where $S_C$ denotes the non-equilibrium
$S$-matrix defined on the Keldysh contour $C$,
\begin{eqnarray}
S_C = T_C \exp \Big\{ -i \gamma \int_C \, dt' \,
[ \psi^\dag(0,t') c(0,t') \nonumber \\  +
c^\dag(0,t') \psi(0,t') ] \Big\} \, .
\end{eqnarray}
$T_C$ is the contour ordering operation. Performing the second
order perturbative expansion in the tunnelling amplitude one
obtains the usual expression for the current \cite{mahan},
\begin{widetext}
\begin{eqnarray}                                     \label{tok}
j(t) &=& \frac{\gamma^2}{2} \int \, dt' \, \Big[ G^{--}_0(t'-t)
g^{--}_0(t,t')
 - G^{--}_0(t-t') g^{--}_0(t',t) \\ \nonumber
&+& G^{+-}_0(t'-t) g^{-+}_0(t,t') - G^{-+}_0(t-t') g^{+-}_0(t',t)
\Big] \, .
\end{eqnarray}
\end{widetext}
We are again using the local Keldysh Green's functions defined as
in Eq.~(\ref{Gfungendefine}). The Green's functions on the
opposite side of the junction, denoted by $G^{ij}_0(t)$, are
defined in a similar way with the change $\psi \rightarrow c$ and
depend, contrary to the situation on the left side of the
junction, only on the time differences. Taking this and the
factorisation relation (\ref{factorisation}) into account we
obtain an expression for the current similar to
Eq.~(\ref{fullnomega}), \widetext
\begin{eqnarray}                                 \label{tokII}
 j(t) = \gamma^2 \int d t' {\cal G}(t') \exp \Big( -i 2 \sin[\Omega
 t'/2] |z| \sin[\Omega t + \Phi] \Big) \, ,
\end{eqnarray}
the function ${\cal G}(t)$ being
\begin{eqnarray}                                 \label{Fcaldefinition}
 {\cal G}(t) = G^{--}_0(-t) g^{--}_0(t)
 - G^{--}_0(t) g^{--}_0(-t)
+ G^{+-}_0(-t) g^{-+}_0(t) - G^{-+}_0(t) g^{+-}_0(-t) \, .
\end{eqnarray}
It can be shown, e.~g. by doing a perturbative expansion in $|z|$, that
the time-ordered part of this function does not contribute to the
tunnelling current and can be dropped.

Again, we expect that it is hardly possible to measure the time
evolution of $j(t)$ directly. A more easily accessible quantity
is the current averaged over one (or more) period of the
oscillating perturbation. Performing the average in the same way
as for the TED we obtain the following general result,
\begin{eqnarray}                             \label{generalresult}
 j = \frac{\gamma^2}{2 \pi} \sum_n J_n^2[|z|] \int d \omega (
 G^{+-}_0(\omega) g^{-+}_0(\omega + \Omega n) - G^{-+}_0(\omega + \Omega n)
 g^{+-}_0(\omega) ) \, .
\end{eqnarray}
\narrowtext
We stress again that we have chosen the specific time-dependence
of the perturbation, (\ref{harmonic}), for the sake of clarity.
A similar formula will hold for any periodic
perturbation (with Bessel functions substituted by
Fourier coefficients relevant for the given perturbation.)

\section{Applications and conclusions} \label{sectionapplications}
The junction is biased by a finite voltage $V$. Setting the
chemical potential in the host to zero and the chemical potential
of the lead to $-V$, we can immediately write down the Green's
functions entering Eq.~(\ref{generalresult}),
\begin{eqnarray}                           \label{1stapplGfs}
 G^{+-}_0(\omega) &=& - i \Theta(\omega+V) \nu_c \, , \\
 G^{-+}_0(\omega) &=&   i \Theta(-\omega-V) \nu_c \, , \nonumber \\
 g^{+-}_0(\omega) &=& - i \Theta(\omega) \nu_\psi(\omega) \, , \nonumber \\
 g^{-+}_0(\omega) &=&   i \Theta(-\omega) \nu_\psi(\omega) \, , \nonumber
\end{eqnarray}
where $\nu_c$ is a constant density of states on the right
electrode and $\nu_\psi(\omega)$ is the LDOS in the host which, in
general, does depend on the energy $\omega$ in the relevant energy
range (set by $V$ and $\Omega$).

Let us first neglect this energy dependence, i.~e. take
$\nu_\psi(\omega)= \nu_\psi^{(0)}$. Then substituting the above
Green's functions into (\ref{generalresult}) we find that because
of the sum rule,
\begin{equation} \label{sum}
\sum\limits_{n=-\infty}^{+\infty}J_n^2[|z|]=1\, ,
\end{equation}
the usual linear current-voltage characteristics \cite{mahan} is
\emph{restored}:
\begin{equation} \label{linear}
j=\gamma^2 \nu_c \nu_\psi^{(0)} V/ 2\pi
\end{equation}
It is remarkable that while the dynamic scatterer profoundly
affects the TED, all these contributions completely cancel out
(for a constant LDOS) in the total current. (A sum rule similar to
(\ref{sum}) and hence the result (\ref{linear}) can be shown to
hold for arbitrary periodic perturbation.) Thus, as long as the
energy dependence of the LDOS can be neglected, no trace of the
dynamic perturbation can be seen in the current-voltage
characteristics. The situation changes dramatically when such
energy dependence is taken into account.

It is instructive to start with a model case, when the host metal
is uncorrelated but the particle-hole symmetry is slightly violated,
i.~e. the LDOS is weakly energy-dependent in vicinity of
the Fermi energy
\begin{eqnarray}
\nu_\psi(\epsilon) = \nu_\psi^{(0)} + \nu_\psi^{(1)} \epsilon\,,
\end{eqnarray}
where $\nu_\psi^{(1)}$ is a constant (the measure of the
particle-hole symmetry violation).
For the current we then obtain
\begin{eqnarray}                   \label{nuresult}
 j = \frac{\gamma^2}{2 \pi} \Big(  \nu_c \nu_\psi^{(0)} V \Big[1-
 \frac{\nu_\psi^{(1)}
 V}{2 \nu_\psi^{(0)}} \Big] +  \nu_c \nu_\psi^{(1)} \Big[ \Omega^2 \sum_1^{|V/\Omega|}
 n^2 J_n^2[|z|] + V \Omega \sum_{|V/\Omega|+1}^{\infty} n J_n^2[|z|]
 \Big] \Big) \, .
\end{eqnarray}
Therefore, as soon as the particle-hole symmetry is violated,
the current becomes strongly influenced by the source of oscillations.
Notice that it is essential to have a finite voltage applied
across the junction.
As is the case for the TED, Eq.~(\ref{nuresult})
can be thought of as a
sum of contributions from different harmonics, that are generated
by the perturbation $U(t)$. The prefactors $J_n^2[|z|]$ govern the
amplitude of $n$-th harmonic. The argument $|z|$ of the Bessel
functions sets a characteristic energy scale and effectively cuts
off the sum over $n$ because $J_n(|z|)$ as a function of $n$
decreases exponentially for $n > |z|$. An important consequence is
that for $|z|<1$ the perturbation is not strong enough to produce
higher harmonics and according to Eq.~(\ref{nuresult}) the linear
behaviour of $j(V)$ is restored with an $\Omega$ dependent
conductance.

\begin{figure}
\includegraphics[scale=0.35]{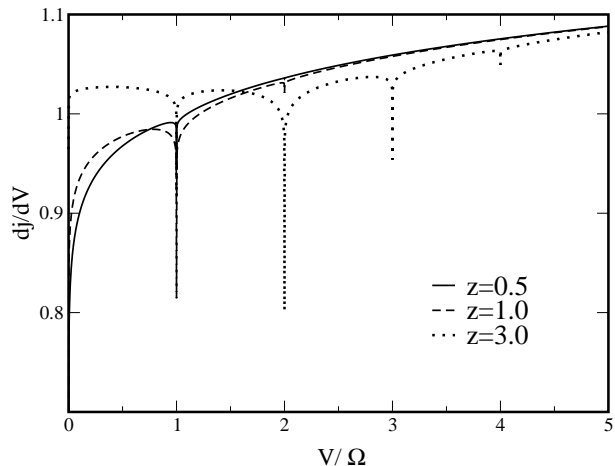}
\caption[]{\label{plot2} Voltage dependence of the differential
conductance for a weakly interacting LL ($g=0.9$)
contacted by a non-interacting metallic
lead. $dj/dV$ is normalised to $g \gamma^2\nu_c C_\psi/2 \pi$. For
a non-interacting system the plot would be a straight line
$dj/dV=1$.}
\end{figure}

In real systems, such as SWNTs, the main reason for the energy
dependence of the LDOS is, of course, correlations. In LLs, the
LDOS is strongly energy-dependent, see Eq.~(\ref{LLDOS}).
Combining Eqs.~(\ref{generalresult}), (\ref{1stapplGfs}) and
(\ref{LLDOS}) we obtain the expression for the current between the
dynamically perturbed LL and a non-interacting lead:
\begin{eqnarray}                            \label{LLresult}
j = g \frac{\gamma^2}{2 \pi} \nu_c C_\psi
\Big\{ J_0^2[|z|]\mbox{sgn}(V) |V|^{1/g}
+ \sum_{n=1}^{\infty} J_n^2[|z|] \Big( \mbox{sgn}(V-\Omega n) |V-\Omega
n|^{1/g} + \mbox{sgn}(V+\Omega n) |V+\Omega n|^{1/g} \Big) \Big\} \, .
\end{eqnarray}
Formula (\ref{LLresult}) is the main result of this paper. As
expected from the previous discussion, turning off the
interactions, $g=1$, restores the linear current-voltage
characteristics expected for a tunnelling junction between two
uncorrelated metallic leads. For the interacting case we predict
frequency and amplitude dependent current across the junction.
Since the SWNTs oscillate at very high frequencies (several
$10^{10}$ Hz which corresponds to $\sim 10^{-4}$ eV, see e.~g.
Ref.~\cite{deheer}) we would expect that in the transport
experiments on oscillating SWNTs even the lowest harmonics should
be observable. The most interesting features arise in the
voltage dependence of the differential conductance $dj/dV$, see
Fig.~\ref{plot2}, where it is plotted for three different
perturbation strengths. At the onset of every harmonic,
$V/\Omega=$ integer there are pronounced dips at the
positions of additional sidebands in the TED.

Another important setup is when a LL is coupled via a tunnelling
junction to a semiconducting electrode (without the bias voltage).
In this case one of the Green's function is identically zero,
$G^{-+}_0(\omega)=0$ as the conductance band of the semiconductor
(SC) is nearly empty. The other off-diagonal Green's function
contains the SC gap $\Delta$, and a constant $\nu_s$ directly
connected with the LDOS in the SC in vicinity of the conductance
band bottom \footnote{Strictly speaking one has to set
$\nu_s(\omega) = \mbox{const}\, \omega/\sqrt{\omega^2 -
\Delta^2}$, but since this energy dependence does not
qualitatively change the results, so we drop it.},
\begin{eqnarray}                            \label{SCGreen}
 G^{+-}_0(\omega) = - i \nu_s \Theta(\omega-\Delta) \, .
\end{eqnarray}
Substituting this into Eq.~(\ref{generalresult}) results in
\begin{eqnarray}                              \label{SCresult}
 j = g \frac{\gamma^2}{2 \pi} \nu_s C_\psi \sum_{n> \Delta/\Omega}^{\infty}
 J_n^2[|z|] (\Omega n - \Delta)^{1/g} \, .
\end{eqnarray}
Contrary to a metallic junction, $j$ is nonzero even in the
non-interacting case. As in the previous setups, the current is
constituted by contributions from different harmonics. However,
since the energy of a given harmonic has to overcome the
SC gap to contribute to the current, the sum starts at
$n=\Delta/\Omega$ (we assume $\Omega$ to be positive). The
tunnelling current shows a very sharp threshold at
$|z|=\Delta/\Omega$ because for $|z|<\Delta/\Omega$ the
contribution of the sum on the right-hand-side
of Eq.~(\ref{SCresult}) is
exponentially small. Of course, Eq.~(\ref{SCresult}) gives the
current immediately after the perturbation was switched on since the
charging effects across the junction would tend to suppress it.
Such process gives rise to a finite potential difference across
the junction which can be detected. As it is related to
the frequency $\Omega$ and amplitude of the oscillations, such
hetero-junction might be employed as a detector.

In conclusion, we analysed the influence of a time-dependent
forward scattering perturbation on the transport properties of
correlated one-dimensional electron systems. We derived the
\emph{exact} Green's functions of the system in the presence of
the dynamic perturbation. This knowledge allowed us to establish
the relation between the total TED of the electrons in the case
with and without the perturbation. In the case of a periodic
perturbation with the frequency $\Omega$ the TED
develops sidebands at energies $E_F + n \Omega$, $n$ being an
integer. The precise energy dependence of the sidebands coincides
with the behaviour of the local density of states in the
particular interacting system without the perturbation. We further
developed this theory to calculate the current-voltage
characteristics of a tunnelling junction between an interacting
electron system and a metallic or semiconducting electrode and
found that the transport is strongly affected by the additional
dynamic perturbation.

\acknowledgements We would like to thank H. Grabert for a valuable discussion.
This work was partly supported by the EPSRC of
the UK under grants GR/N19359 and GR/R70309.

\bibliography{prb2}
\end{document}